\documentclass[twocolumn,times]{aastex63}
\usepackage{hyperref}
\hypersetup{
    hidelinks,
    colorlinks=true,
    allcolors=black,
    pdfstartview=Fit,
    breaklinks=true,
    linkcolor=blue,
    urlcolor=blue,
    citecolor=blue}

\begin{document}

\title{GRB 211211A: a Prolonged Central Engine under a Strong Magnetic Field Environment}

\correspondingauthor{He Gao}
\email{gaohe@bnu.edu.cn}
\correspondingauthor{Wei-Hua Lei}
\email{leiwh@hust.edu.cn}

\author[0000-0002-3100-6558]{He Gao}
\affiliation{Department of Astronomy ,
Beijing Normal University, Beijing, People's Republic of China}

\author[0000-0003-3440-1526]{Wei-Hua Lei}
\affiliation{Department of Astronomy , School of Physics,
Huazhong University of Science and Technology, Wuhan, Hubei 430074, China}

\author{Zi-Pei Zhu}
\affiliation{Department of Astronomy , School of Physics,
Huazhong University of Science and Technology, Wuhan, Hubei 430074, China}
\affiliation{Key Laboratory of Space Astronomy and Technology, National Astronomical Observatories, Chinese Academy of Sciences, Beijing 100012, China}

\begin{abstract}
\noindent
Recently, a kilonova-associated gamma-ray burst (GRB 211211A), whose light curve consists of a precursor ($\sim$0.2 s), a hard spiky emission ($\sim$10 s), and a soft long extended emission ($\sim$40 s), has attracted great attention. Kilonova association could prove its merger origin, while the detection of the precursor can be used to infer at least one highly magnetized neutron star (NS) being involved in the merger. In this case, a strong magnetic flux $\Phi$ is expected to surround the central engine of GRB 211211A. Here we suggest that when $\Phi$ is large enough, the accretion flow could be halted far from the innermost stable radius, which will significantly prolong the lifetime of the accretion process, and so the GRB duration. For example, we show that as long as the central black hole (BH) is surrounded by a strong magnetic flux $\Phi\sim 10^{29}\rm cm^{2}G$, an accretion flow with $\dot{M}_{\rm ini} \simeq 0.1 M_\odot s^{-1}$ could be halted at 40 times the gravitational radius and be slowly transferred into the black hole on the order of $\sim$10 s, which naturally explains the duration of hard spiky emission. After most of the disk mass has been accreted onto the BH, the inflow rate will be reduced, so a long and soft extended emission is expected when a new balance between the magnetic field and the accretion current is reconstructed at a larger radius. Our results further support that the special behavior of GRB 211211A is mainly due to the strong magnetic field of its progenitor stars.
\end{abstract}

\keywords{\href{http://astrothesaurus.org/uat/629}{Gamma-ray bursts (629)}}

\section{Introduction}

Phenomenologically, gamma-ray bursts (GRBs) are classified into two categories: the long-duration, soft-spectrum class (LGRBs) and the short-duration, hard-spectrum class (SGRBs), based on the bimodal distribution of GRBs in the duration–hardness diagram \citep{1993ApJ...413L.101K}. The boundary between the two classes is vague and instrument dependent \citep{2013ApJ...763...15Q}. Traditionally, an observer-frame duration $T_{90}$ $\sim$ 2 s is taken to be the separation line: bursts with $T_{90}$ $>$ 2 s are long and bursts with $T_{90}$ $<$ 2 s are “short.”

Different types of progenitors are invoked in the theory for these two different classes, i.e., core collapse from a Wolf-Rayet star for LGRBs \citep{1993ApJ...405..273W,1998ApJ...494L..45P,1999ApJ...524..262M,2006ARA&A..44..507W} and mergers of two compact stellar objects (NS–NS and NS–BH systems) for SGRBs \citep{1986ApJ...308L..43P,1989Natur.340..126E,1991AcA....41..257P,1992ApJ...395L..83N}. Many important observations seem to support such interpretations for the progenitor; for instance, most host galaxies of LGRBs are irregular, star-forming galaxies, with a few being spiral galaxies with active star formation \citep{2006Natur.441..463F}, while the majority of host galaxies of short GRBs are elliptical or early type \citep{2005Natur.437..851G}. On the other hand, the offset of the SGRB location with respect to the center of their host galaxy is systemically larger than that of the LGRBs \citep{2014ARA&A..52...43B}. For the collapsar scenario, the most direct evidence is that a handful of LGRBs are firmly associated with Type Ib/c supernovae (SNe) \citep{2006ARA&A..44..507W}. For the merger scenario, the smoking-gun evidence was established by the association between the gravitational-wave-detected binary NS (BNS) merger, GW170817, and the weak short-duration GRB 170817A \citep{2017ApJ...848L..13A}. 

Most recently, the peculiar LGRB 211211A, detected by Fermi/GBM \citep{2021GCN.31210....1M}, Swift/BAT \citep{2021GCN.31202....1D} and Insight-HXMT/HE \citep{2021GCN.31236....1Z}, has severely challenged this clean dichotomy of the two populations. The total duration of this burst is $51.37 \pm 0.80 $s in BAT ($\sim$34.3 s in GBM), whose light curve contains three emission episodes \citep{2022arXiv220502186X}: a precursor with a duration of $\sim 0.2$ s, a $\sim 10$ s spiky hard main emission (ME), and a soft long extended emission (EE) up to $\gtrsim $50 s \citep{2021GCN.31210....1M}. This source has attracted great attention because it phenomenologically definitely belongs to the long-duration category (even without counting the EE part), but many obvious evidence links it to a compact object merger \citep{2022arXiv220410864R,2022arXiv220412771Y,2022arXiv220502186X}: (1) Despite its promising proximity, surprisingly no SN was observed to accompany the GRB down to very deep detection limits, yet an associated kilonova was discovered based on the detailed analysis of observation data from multiple optical-NIR telescopes; (2) the physical offset between the burst and the nucleus of the host galaxy is more consistent with the known offsets of SGRBs; and (3) the spectral lag of the ME is consistent with the known values of SGRBs, but obviously deviate from LGRBs in the spectral lag - $L_{\rm iso}$ diagram. The biggest challenge this burst poses to theorists is: In the context of the merger scenario, how does the hard spiky emission last for $\sim 10$ s? 

\begin{figure}[http]
\includegraphics[width=\columnwidth]{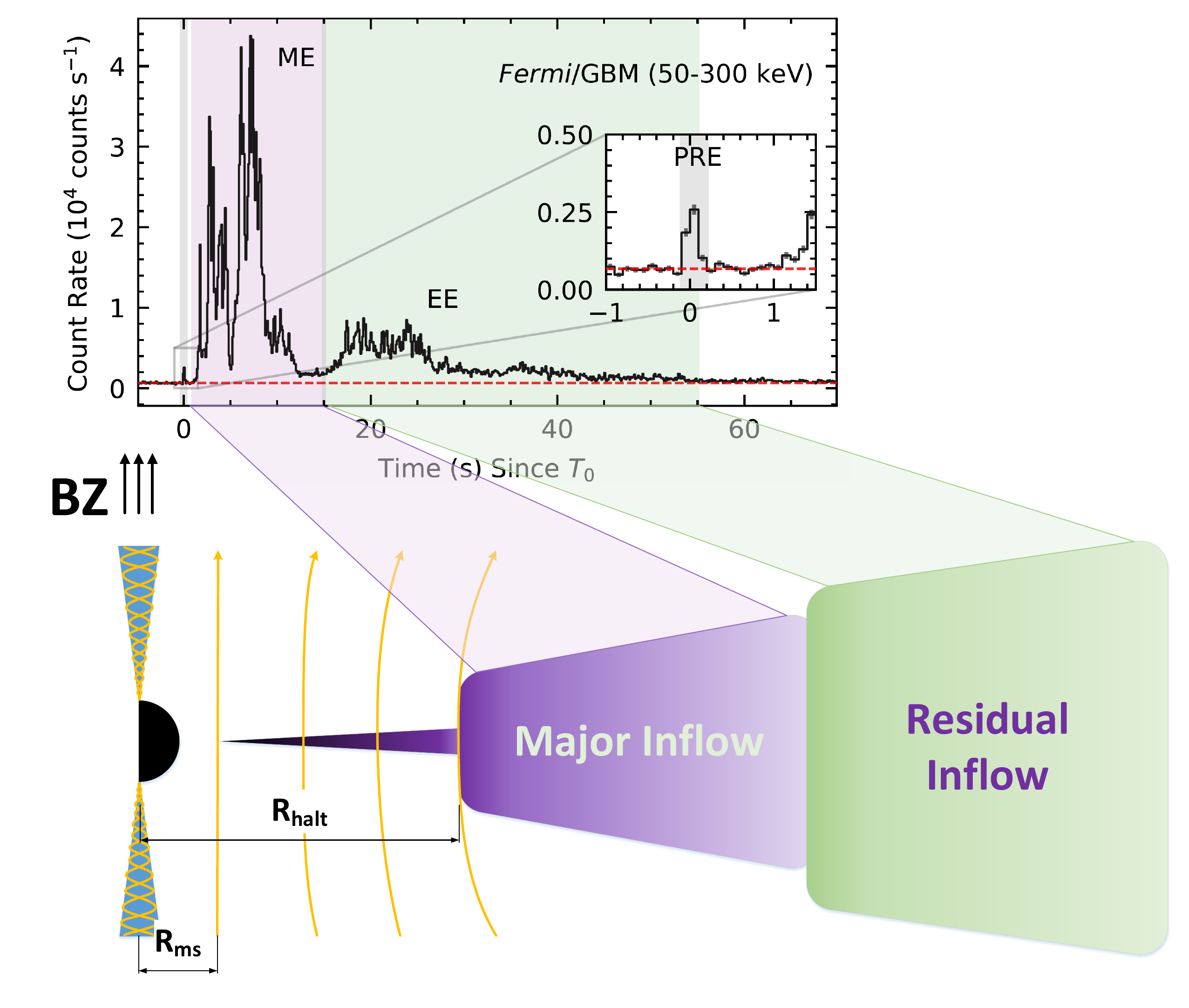}
\caption {Illustration of our model. With the magnetic barrier effect, the accretion of the major inflow corresponds to the hard spiky ME of GRB 211211A, and the accretion of the residual inflow corresponds to the soft EE of GRB 211211A. The GRB jet is powered by the Blandford–Znajek (BZ) mechanism, which extracts the rotational energy of the Kerr BH through a large-scale magnetic field. Here $R_{\rm halt}$ marks the radius where the radial magnetic force can give support against the gravitational force. $R_{\rm ms}$ marks the radius of the marginally stable orbit. The light-curve picture of GRB 211211A is adopted from \cite{2022arXiv220502186X}.}
\label{cartoon}
\end{figure}

Based on the special properties of the precursor (e.g., the waiting time between it and the spiky main emission is $\sim$ 1 s, which is about the same as the time interval between GW 170817 and GRB 170817A), especially its claimed $\sim$22 Hz quasi-periodic oscillations, it is proposed that the progenitor system of GRB 211211A very likely contains a magnetar, and the seismic aftershocks and low-frequency torsional modes may explain the underlying oscillations once the precursor results from the resonant shattering of the magnetar \citep{2022arXiv220511112S,2022arXiv220502186X}. In this case, the strong seed magnetic field of $\sim10^{14}-10^{15}$ G at the surface of the magnetar could leave behind a strong magnetic flux $\Phi$ surrounding the central engine of GRBs \citep{2014PhRvD..90d1502K}. It has long been proposed that due to the magnetic barrier effect, radial angular momentum transfer may significantly prolong the lifetime of the accretion process, and so the GRB duration \citep{2006MNRAS.370L..61P, 2012ApJ...760...63L}. In this work, we further refine the magnetic barrier model (see Figure \ref{cartoon} for a cartoon picture) and search for a reasonable parameter space to interpret the observations of GRB 211211A. 

\section{Magnetic Barrier Model}

We consider a compact binary merger with at least one highly magnetized NS being involved. During the merger, the magnetized NS would be disrupted due to tidal force and destroyed on collision. Here we focus on the case where the merger remnant is a BH surrounded by an accretion torus.\footnote{For NS-NS mergers, the product could be either a BH or a supramassive NS, depending on the total mass of the NS-NS system and the NS equation of state  \citep{lasky14,rosswog14,gao16}. The merger product for GRB 211211A is very likely a BH, otherwise if the product is a supramassive NS, the associated kilonova should be much brighter (see \citealt{yu13,metzger14}, etc.)}. In this case, the magnetization of the NS would be preserved by the debris, serving as magnetic filed seeds. MHD instabilities would develop and amplify the magnetic fields \citep{rezzolla11,ciolfi17}. The GRB's central engine is thus a spinning black hole with mass $M_\bullet$ and angular momentum $J_\bullet$, surrounded by a strong magnetic flux $\Phi$. Before encountering the magnetic barrier, the initial mass flow rate is $\dot{M}_{\rm ini}$. The existence of a strong magnetic field could have a significant impact on the accretion process. First, the accretion flow will be halted by a magnetic barrier at some radius $R_{\rm halt}$. This occurs when the radial magnetic force can support against the gravitational force \citep{2006MNRAS.370L..61P,2012ApJ...760...63L,2018MNRAS.475..266Z},
\begin{eqnarray}
\frac{2 B_{\rm R} B_{\rm Z}}{4\pi} \simeq \frac{G M_\bullet \Sigma}{R_{\rm halt}^2},
\end{eqnarray}
where $\Sigma$ is the surface density at $R_{\rm halt}$, which can be given by
\begin{equation}
\Sigma = \frac{\dot{M}_{\rm ini}}{2\pi R_{\rm halt} \epsilon_{\rm i} v_{\rm ff}}.
\end{equation}
Here the flow radial velocity, when the magnetic field is just in contact with the accretion flow, is assumed to be a fraction $\epsilon_{\rm i}$ of the freefall velocity $v_{\rm ff}=\sqrt{G M_\bullet/r}$. It is reasonable to adopt $\epsilon_{\rm i} = 10^{-3} - 10^{-2}$ \citep{2006MNRAS.370L..61P,2012ApJ...760...63L,2018MNRAS.475..266Z}. 

We thus have 
\begin{eqnarray}
r_{\rm halt} = \frac{R_{\rm halt}}{R_{\rm g}} = 91.0 \epsilon_{\rm i,-2}^{2/3}  m_3^{-4/3} \dot{m}_{\rm ini,-1}^{-2/3} \Phi_{30}^{4/3} ,
\label{eq:rm}
\end{eqnarray}
where $R_{\rm g} = G M_\bullet/c^2$, $\Phi_{30}= \Phi/(10^{30} \rm cm^2 G)$, $m_3=M_\bullet/3M_\odot$, and $\dot{m}_{\rm ini,-1}=\dot{M}/0.1M_\odot \rm s^{-1}$. Here we assume $B_{\rm R}\simeq B_{\rm Z}=B$, and the magnetic flux is connected with the magnetic field as $\Phi\simeq \pi R^2 B$. 

As mass accumulated at $R_{\rm halt}$, the accretion will restart with a much lower radial velocity $\epsilon_{\rm m} v_{\rm ff} \ll  \epsilon_{\rm i} v_{\rm ff}$, due to the magnetic tension. The accretion time can be estimated as
\begin{eqnarray}
t_{\rm acc} = \frac{R_{\rm halt}}{\epsilon_{\rm m} v_{\rm ff}} \simeq 13 {\rm s} \frac{\epsilon_{\rm i,-2}}{\epsilon_{\rm m,-3}} m_3^{-1} \dot{m}_{\rm ini,-1}^{-1} \Phi_{30}^2.
\end{eqnarray}
Therefore, a larger magnetic flux leads to a larger halting radius $R_{\rm halt}$ and a longer accretion duration $t_{\rm acc}$. 

The initial magnetic filed threading the BH horizon can be estimated as (assuming the magnetic field to be a uniform field)
\begin{eqnarray}
B_{\rm ini} \sim 1.9 \times 10^{14} {\rm G} \ \  m_3^{2/3} \dot{m}_{\rm ini,-1}^{4/3} \epsilon_{\rm i,-2}^{-4/3} \Phi_{30}^{-5/3}.
\end{eqnarray}
The $B$ field will increase as the flow pushes the magnetic flux to the radius of the marginally stable orbit $R_{\rm ms}$. The $B$ field at $R_{\rm ms}$ is 
\begin{eqnarray}
B_{\rm ms}=4.5 \times 10^{16} {\rm G} \ \  m_3^{-2} \Phi_{30} (r_{\rm ms}/6)^{-2},
\end{eqnarray}
where the radius of the marginally stable orbit is expressed as \citep{1972ApJ...178..347B}
\begin{equation}
r_{\rm ms}= \frac{R_{\rm ms}}{R_{\rm g}} = 3+Z_2 - [(3-Z_1)(3+Z_1+2Z_2)]^{1/2}, 
\end{equation}
for $0\leq a_{\bullet} \leq 1$, where $a_\bullet$ is the BH spin parameter defined as $a_\bullet=J_\bullet c/G M_\bullet^2$, $Z_1 \equiv 1+(1-a_\bullet^2)^{1/3} [(1+a_\bullet)^{1/3}+(1-a_\bullet)^{1/3}]$ and $Z_2\equiv (3a_\bullet^2+Z_1^2)^{1/2}$.  

In this model, the GRB prompt emission can be powered by the BZ \citep{1977MNRAS.179..433B} mechanism, in which the spin energy of the BH is extracted via the open field lines penetrating the event horizon. The BZ jet power could be estimated as \citep{2000PhR...325...83L,2000PhRvD..61h4016L,2002MNRAS.335..655W,2005ApJ...630L...5M,Lei05,2011ApJ...740L..27L,2013ApJ...765..125L,2015ApJS..218...12L}
\begin{eqnarray}
\dot{E}_{\rm B}=1.7 \times 10^{50}  a_{\bullet}^2 m_{\bullet}^2
B_{\bullet,15}^2 F(a_{\bullet}) \ {\rm erg \ s^{-1}} \\ \nonumber
\simeq 1.1 \times 10^{50}  a_{\bullet}^2 m_{\bullet}^2
B_{\bullet,15}^2 \ {\rm erg \ s^{-1}},
\label{BZ}
\end{eqnarray}
where $B_{\bullet,15}=B_{\bullet}/10^{15} {\rm G}$ and $F(a_\bullet)=[(1+q^2)/q^2][(q+1/q) \arctan q-1]$. Here $q= a_{\bullet} /(1+\sqrt{1-a^2_{\bullet}})$, and $2/3\leq F(a_{\bullet}) \leq \pi-2$ for $0\leq a_{\bullet} \leq 1$. 

The prompt emission would be carried out in two parts. The first part is within the timescale of  $t_{\rm acc}$. In this part, most of the accretion flow will fall into the BH. The accretion rate would normally be higher than the igniting accretion rates $\dot{m}_{\rm ign}$ for neutrino-emitting reactions, especially considering that a strong magnetic field could effectively reduce $\dot{m}_{\rm ign}$ \citep{2009ApJ...700.1970L}, so the hyperaccreting disk would be neutrino dominated. In this case, the baryon-loading rate for the BZ-driven jet could be estimated as \citep{2013ApJ...765..125L}
\begin{eqnarray}
\dot{M}_{\rm j,BZ} & \simeq & 3.5 \times 10^{-7} A^{0.58} B^{-0.83} f_{\rm p,-1}^{-0.5} \theta_{\rm j,-1} \theta_{\rm B,-2}^{-1} \nonumber \\
& & \times  \alpha_{-1}^{0.38} \epsilon_{-1}^{0.83} \dot{m}_{-1}^{0.83} m_3^{-0.55}  r_{z,11}^{0.5} \ M_{\sun} {\rm s}^{-1},
\label{Eq_mdotj_BZ}
\end{eqnarray}
where $A=1-2r^{-1}+a_\bullet^2 r^{-2}$ and $B=1-3r^{-1}+2 a_\bullet r^{-3/2}$ are the general relativistic correction factors for a thin accretion disk around a Kerr BH \citep{1995ApJ...450..508R}, $f_{\rm p}$ is the fraction of protons, $\theta_{\rm j}$ is the jet-opening angle, $\alpha$ is the dimensionless viscosity parameter \citep{1973A&A....24..337S}, $r_{\rm z, 11}$ is the distance from the BH in the jet direction normalized to $10^{11}$cm. Protons with an ejection direction larger than $\theta_{\rm B}$ with respect to the field lines would be blocked due to the existence of a strong magnetic field. Apparently, baryons from the disk will be suppressed by the strong magnetic field lines. The maximum available energy per baryon in the BZ jet can be denoted by parameter $\mu_0 $ as
\begin{equation}
\mu_0 \equiv \frac{\dot{E}}{\dot{M}_{\rm j,BZ} c^2} = \frac{\dot{E}_{\rm m}+\dot{E}_{\rm B}}{\dot{M}_{\rm j,BZ} c^2} = \eta (1+\sigma_0),
\label{eq:mu}
\end{equation}
where $\dot{E}_{\rm m} = \dot{E}_{\nu \bar{\nu}}+ \dot{M}_{\rm j,BZ} c^2$ and $\sigma_0 = \dot E_{\rm B}/\dot E_{\rm m}$, $\eta=\dot{E}_{\nu \bar{\nu}} /(\dot{M}_{\rm j, BZ} c^2)$, where $\dot{E}_{\nu \bar{\nu}}$ is the neutrino annihilation power \citep{2015ApJS..218...12L,2017NewAR..79....1L,2017ApJ...849...47L}
\begin{eqnarray}
\dot{E}_{\nu \bar{\nu}}  \simeq  \dot{E}_{\nu \bar{\nu}, \rm ign}  \left(\frac{\dot{m}}{\dot{m}_{\rm ign}} \right)^{2.23 }    ,
\label{eq_Evv}
\end{eqnarray}
for $\dot{m}>\dot{m}_{\rm ign}$, where 
\begin{eqnarray}
\dot{E}_{\nu \bar{\nu}, \rm ign}=10^{(48.0+0.15 a_\bullet)}  m_3^{\log(\dot{m}/\dot{m}_{\rm ign}) -3.3} {\rm erg \ s^{-1}}.
\end{eqnarray}
The magnetic dissipation and acceleration dynamics of the jet are quite uncertain. We take $\Gamma_{\rm min}=\mathbf{max}(\mu_{0}^{1/3},\eta)$ and $\Gamma_{\rm max} =  \mu_0$, which correspond to the beginning and the end of the slow acceleration phase in a hybrid outflow \citep{2015ApJ...801..103G}. The jet will reach a terminating Lorentz factor $ \Gamma_{\rm min} < \Gamma < \Gamma_{\rm max}$. For typical parameters, relatively large values for $\Gamma$ and $\sigma$ are expected. So, the first part of emission would be hard spiky emission with duration $t_{\rm acc}$. 

Considering both the accretion and BZ processes, the evolution equations of BH are given by
\begin{equation}
\frac{dM_\bullet c^2}{dt} = \dot{M} c^2 E_{\rm ms} - \dot{E}_{\rm B},
\label{dMbz}
\end{equation}

\begin{equation}
\frac{dJ_\bullet}{dt} = \dot{M} L_{\rm ms} - T_{\rm B}
\label{dJbz}
\end{equation}
where $E_{\rm ms}$ and $L_{\rm ms}$ are the specific energy and the specific momentum corresponding to the innermost radius $r_{\rm ms}$ of the disk, which are defined in \cite{1973blho.conf..343N} as
$E_{\rm ms} = (4\sqrt{ r_{\rm ms} }-3a_{\bullet}) /(\sqrt{3} r_{\rm ms})$, $L_{\rm ms} = (G M_\bullet/c) (2 (3 \sqrt{r_{\rm ms}} -2 a_\bullet) )/(\sqrt{3} \sqrt{r_{\rm ms}} )$. The BZ torque applied on the BH is $T_{\rm B}=\dot{E}_{\rm B}/(0.5 \Omega_\bullet)$ and the angular velocity of the BH horizon is $\Omega_\bullet=q c/(2R_{\rm g})$.

After the first part, because most of the disk mass has been accreted onto the BH, the inflow rate of the remaining mass should be quite low. A majority of the magnetic flux would quickly diffuse out \citep{2012ApJ...760...63L,2018MNRAS.475..266Z}. The balance between the magnetic field and the accretion current will be reconstructed at a new and further radius. Nevertheless, because the magnetic field and accretion rate decrease at the same time, it is difficult to reach the ignition threshold of neutrino-emitting reactions. The disk will thus be advection dominated, which has a strong disk wind driven by a positive Bernoulli constant \citep{1994ApJ...428L..13N,2018MNRAS.477.2173S}. For typical parameters, relatively small values for $\Gamma$ and $\sigma$ are expected. So, the second part of emission would be longer and softer than the first part. Depending on the detector's sensitivity, the second emission episodes may not be detected or may be detected as an extended emission phase (see detailed modeling for this part in \citealt{2012ApJ...760...63L}).  

For the merger events involving NSs with an ordinary magnetic field, the magnetic flux around the central BH may not be strong enough to block the accretion flow resistance beyond the radius of the innermost stable orbit. Therefore, a short-duration ME will be generated. For such events (including GRB 211211A–like events), whether the ME would be followed by an EE depends on the nature of the late accretion \citep{2012ApJ...760...63L}. 

\section{Application to GRB 211211A}

The prompt emission of GRB 211211A could be divided into three episodes: a precursor with duration of $\sim$0.2 s, a $\sim$10 s spiky hard ME, and a soft long EE up to $\gtrsim$50 s. Here we show that the ME and EE parts could be well interpreted within the magnetic barrier model, as long as the central engine of GRB 211211A contained a newly formed BH with mass $M_\bullet=\sim 2.5 M_\odot$, initial magnetic flux of $\Phi_{30}\simeq 0.2$, and a relatively low initial BH spin $a_\bullet = 0.1$ (low spin is due to the possible rotation braking caused by strong magnetic field). The initial accretion flow rate is taken as $\dot{m}_{\rm ini} \simeq 0.1$, which is a typical value for regular SGRBs. 

Firstly, from Equation (\ref{eq:rm}), the accretion flow will be halted at a radius $R\simeq 40R_{\rm g}$ due to the magnetic barrier. The accretion timescale is $t_{\rm acc}\simeq 10s$ for $\epsilon_{\rm i}=0.05$ and $\epsilon_{\rm m}=3\times 10^{-4}$, which explains the duration of ME. 

Secondly, during the first 10 s, the average magnetic filed threading BH would be $\sim 10^{16}$ G. From Equation (8), the luminosity of BZ jet would be $\sim 10^{51} {\rm erg\ s^{-1}}$, which is well consistent with the jet-corrected energies $E_{\gamma}+E_{K}$ for ME \citep{2022arXiv220502186X}. 
 
Third, for typical values of $A=0.8$, $B=0.7$, $\alpha=0.01$, $f_{\rm p}=0.1$, $\theta_{\rm B}=0.01$ and $r_{\rm z,11}=1$ \cite[][for details]{2017ApJ...849...47L}, from Equation (\ref{eq:mu}), the maximum available energy per baryon in the BZ jet would be $\mu_0 \simeq 3000$. The acceleration behavior of the jet is subject to uncertainties. The jet will reach a terminating Lorentz factor $\Gamma_{\rm min}<\Gamma<\Gamma_{\rm max}$ with the explicit value depending on the detailed dissipation process, where $\Gamma_{\rm min} = \mathbf{max} (\mu_0^{1/3}, \eta)$ and $\Gamma_{\rm max} =\mu_0$. In this case, the Lorentz factor of the jet will be larger than $\eta=400$, which could be the reason why ME consists of many spiky pulses exhibiting little spectral evolution.

Finally, after the first 10 s, the majority of the remnant disk mass (e.g., $\sim 0.2 M_\odot$) is accreted. The inflow rate of the remaining mass is largely reduced, so is the magnetic flux maintaining by the inflow. As shown in the simulation of \cite{2019MNRAS.482.3373F}, $\sim$1\% of disk matter could reach a radius of $\sim2000R_{\rm g}$ keeping it gravitationally bound. We suggest that the fallback accretion of these matter could power the extended emission. Here we take $\dot{m}_{\rm flow}\simeq 0.001$ (1\% of the initial accretion flow rate) and $\Phi_{30} \simeq 0.07$. In this case, the accretion flow will be halted at radius $R\simeq 200R_{\rm g}$ due to the magnetic barrier. The accretion timescale is $t_{\rm acc}\simeq 40s$ for $\epsilon_{\rm i}=0.05$ and $\epsilon_{\rm m}=8\times 10^{-4}$, which explains the duration of EE. Due to the accretion, the BH spin has increased to $a_\bullet = 0.3$. At this stage, the magnetic filed threading the BH would be up to $\sim 10^{15}$ G, so the luminosity of the BZ jet would be $\sim 10^{50} {\rm erg \ s^{-1}}$, which is also consistent with the jet-corrected energies $E_{\gamma}$ for EE \citep{2022arXiv220502186X}.

\section{Conclusion and Discussion}
 
Compared with other EE-SGRBs, GRB 211211A has three particularities: (1) an associated KN has been found, (2) a peculiar precursor has been found, (3) the ME duration prior to EE is in order of 10 s. The KN association could prove its merger origin, while the detection of the precursor can be used to infer at least one highly magnetized NS being involved in the merger. Numerical relativity simulations have already shown that for the merger of binary highly magnetized NSs, the central engine of GRBs would be modeled by a highly magnetized accretion torus \citep{2014PhRvD..90d1502K}. Here we show that such a scenario could naturally explain the long ME duration by invoking magnetic barrier effect. We find that as long as the central BH is surrounded by a strong magnetic flux $\Phi\sim 10^{29}\rm cm^2 G$, an accretion flow with $\dot{m}_{\rm ini} \simeq 0.1$ could be halted at $40R_{\rm g}$ and slowly transfer into the black hole on the order of $\sim$10 s.

According to numerical simulations, the strength and structure of the magnetic field around the BH, as well as the relationship between the magnetic field and the accretion disk, are very uncertain and highly depend on the selection of initial conditions \citep{rezzolla11,ciolfi17}. Some assumptions introduced here may bring some uncertainty to the results. For instance, we assume that the magnetic field is mainly composed of open lines (aligned with the rotation axis of the BH), part of which threads the nascent BH horizon, and other parts are distributed outside the horizon, trying to spread outward in the direction of the disk, so as to push outward the accretion torus to a balancing point ($R_{\rm halt}$).  Here we ignore the spinning effect of the magnetic field lines. For a given balancing point, the spinning effect could extend the accretion timescale. In this case, the balancing point for the ME/EE part of GRB 211211A could be smaller than $40R_{\rm g}$/$200R_{\rm g}$. Moreover, when the spinning effect is considered, part of the accretion gas would be ejected along the magnetic field lines, similar to the propeller mechanism \citep{1975A&A....39..185I}. These outflow materials could increase the baryon-loading rate for the GRB jet, reducing the magnetized factor and thus reducing the terminating Lorentz factor of the jet.

If the closed field lines are nonnegligible, energy and angular momentum will be transferred between the BH and the surrounding disk; such a mechanism is referred to as the magnetic coupling (MC) process. The magnetic coupling between the central spinning BH and their surrounding
accretion disk also plays an important role in GRB central engine \citep{2009ApJ...700.1970L} . Due to the freezing-in condition in the disk, the angular velocity of the closed field lines connecting the BH and the disk is equal to the angular velocity of the disk $\Omega_{\rm D}=((R^3/G M_\bullet)^{1/2} + a_\bullet G M_\bullet/c^3 )^{-1}$. The transfer direction of energy and angular momentum between the BH and the disk is determined by the ratio $\beta=\Omega_{\rm D}/\Omega_\bullet$ of the angular velocity of the disk $\Omega_{\rm D}$ to that of the BH horizon $\Omega_\bullet$. Defining the corotation radius $R_{\rm co}$ as the radius on the disk where the angular velocity of the disk is equal to the BH angular velocity, $\beta=1$. Inside $R_{\rm co}$, energy and angular momentum are transferred by the closed magnetic field lines from the disk into the BH with $\Omega_D>\Omega_\bullet$, while the transfer direction reverses for $R>R_{\rm co}$ with $\Omega_D<\Omega_\bullet$. For $a_\bullet=0.1 (0.3)$ during ME (EE), we have $r_{\rm co}\simeq 11.7 (5.5)<r_{\rm halt}$, i.e., $\beta<1$ in the initial accretion flow. Therefore, for our case of interest, BH rotates fast than the disk, the MC process exerts a torque $T_{\rm MC}=1.3\times 10^{46} m_\bullet^3 B_{\bullet,15}^2 a_\bullet (1+\sqrt{1-a_\bullet^2}) \int_0^{\pi/2} \frac{(1-\beta) \sin^3\theta d\theta}{2-(1-\sqrt{1-a_\bullet^2}) \sin^2\theta} \rm g \ cm^2 s^{-2}$ on the disk, and energy and angular momentum are transferred from the BH into the disk. This process will help halt the flow or even push $r_{\rm halt}$ to a larger radius. When the accretion restarts, the MC torque may result in an even lower accretion rate due to the extra angular momentum (from the BH) to be transferred by the flow. In \cite{2009ApJ...700.1970L}, it is found that the luminosity of neutrino annihilation will be enhanced when the MC process is involved. The effect depends on the ratio $\eta_{Q}=Q_{\rm MC}/Q_{\rm G}$, where $Q_{\rm G} =3G M_\bullet \dot{M} /(8\pi R^3)$ and $Q_{\rm MC} = -T_{\rm MC} \Omega_{\rm D}^\prime /(4\pi R)$ are the contributions due to the gravitational release and the MC process, respectively. For GRB 211211A, we find that $\eta_{\rm Q}\ll 1$ during ME, so the enhancement of neutrino annihilation luminosity due to MC can be ignored. 

On the other hand, the ratio between $\epsilon_i$ and $\epsilon_m$ is essential for estimating the accretion timescale. Some previous studies have applied the magnetic barrier model to interpret the late X-ray flares and extended emission of GRBs, and their results suggest $\epsilon$ to be on the order of $10^{-2}-10^{-3}$ \citep{2006MNRAS.370L..61P,2012ApJ...760...63L}. In this work, we take $\epsilon_i\sim10^{-2}$ and $\epsilon_m\sim10^{-3}$ as fiducial values. Specific numerical simulations are required in the future to justify this assumption.

Besides the peculiar long ME duration and KN association, some other intriguing properties for GRB 211211A have also been proposed. For instance, a significant ($>5\sigma$) transient-like emission in the high energy gamma-rays ($>100$ MeV) was observed by Fermi/LAT starting at $10^3$s after the burst and lasting $\sim2\times10^4$ s \citep{2022arXiv220508566M,2022arXiv220509675Z}. The unusually long duration of the GeV emission might indicate that the GRB jet is expanding in an extremely low-density circumburst medium, consistent with the compact stellar merger scenario, especially when the pulsar wind from the magnetized NS may have created a cavity around the system \citep{2022arXiv220509675Z}. On the other hand, \cite{2022arXiv220505008G} finds that the rapidly evolving spectrum of GRB 211211A can be fitted by purely synchrotron emission within the so-called "marginally fast-cooling regime," inferring that accelerated particles do not cool completely via synchrotron processes within a dynamical timescale. They thus suggest that for a typical bulk Lorentz factor of $\Gamma\sim100$, the magnetic field falls in the range $30-200$ G for a range of $R\sim10^{13}-10^{14}$ cm. It is worth noting that for a magnetically dominated jet, significant magnetic dissipation could occur to distort the magnetic lines, resulting in a reconnection cascade and thus a significant release of the stored magnetic field energy to power the observed GRB prompt emission \citep{zhangyan11,2019ApJ...882..184L}. In this case, a relatively low magnetic field ($10-10^{4}$ G) is expected in the emission region (e.g., a reconnection layer) compared with the undissipated regions in the outflow \citep{uhm14,shao22}, consistent with the results shown in \cite{2022arXiv220505008G}.

Numerical simulations show that the gravitational waveforms of magnetized and unmagnetized NS binaries could be well distinguished as long as the NS magnetic field is strong enough \citep{2009MNRAS.399L.164G}. In the future, multimessenger detections of GRB 211211A-like events could help to diagnose their progenitor system and thus justify our model. Nevertheless, it is worth noting that GRB 211211A-like events could be disguised as a typical LGRBs, once their EEs are too weak to be recognized. LGRBs with $T_{90}\gtrsim$10 s, especially those with clear signatures of merger origin (e.g., small spectral lag, large host galaxy offset, KN association, etc.), should also be carefully studied, which might be help in the estimation of the event rate of merging events involving high-magnetic-field NSs. 

\section*{Acknowledgments}

We thank the anonymous referee for the helpful comments that have helped us improve the presentation of the paper. This work is supported by the National Natural Science Foundation of China (Projects: 12021003, U2038107, U1931203). We acknowledge the science research grants from the China Manned Space Project with Nos. CMS-CSST-2021-A13 and CMS-CSST-2021-B11.

\section*{ORCID iDs}
\noindent
He Gao \href{https://orcid.org/0000-0002-3100-6558}{https://orcid.org/0000-0002-3100-6558}

\noindent
Wei-Hua Lei \href{https://orcid.org/0000-0003-3440-1526}{https://orcid.org/0000-0003-3440-1526}



\begin{thebibliography}{99}

\bibitem[Abbott et al.(2017)]{2017ApJ...848L..13A} Abbott, B.~P., Abbott, R., Abbott, T.~D., et al.\ 2017, \apjl, 848, L13. doi:10.3847/2041-8213/aa920c
\bibitem[Bardeen et al.(1972)]{1972ApJ...178..347B} Bardeen, J.~M., Press, W.~H., \& Teukolsky, S.~A.\ 1972, \apj, 178, 347. doi:10.1086/151796
\bibitem[Berger(2014)]{2014ARA&A..52...43B} Berger, E.\ 2014, \araa, 52, 43. doi:10.1146/annurev-astro-081913-035926
\bibitem[Blandford \& Znajek(1977)]{1977MNRAS.179..433B} Blandford, R.~D. \& Znajek, R.~L.\ 1977, \mnras, 179, 433. doi:10.1093/mnras/179.3.433
\bibitem[Ciolfi et al.(2017)]{ciolfi17} Ciolfi, R., Kastaun, W., Giacomazzo, B., et al.\ 2017, \prd, 95, 063016. doi:10.1103/PhysRevD.95.063016
\bibitem[D'Ai et al.(2021)]{2021GCN.31202....1D} D'Ai, A., Ambrosi, E., D'Elia, V., et al.\ 2021, GRB Coordinates Network, Circular Service, No. 31202, 31202
\bibitem[Eichler et al.(1989)]{1989Natur.340..126E} Eichler, D., Livio, M., Piran, T., et al.\ 1989, \nat, 340, 126. doi:10.1038/340126a0
\bibitem[Fern{\'a}ndez et al.(2019)]{2019MNRAS.482.3373F} Fern{\'a}ndez, R., Tchekhovskoy, A., Quataert, E., et al.\ 2019, \mnras, 482, 3373. doi:10.1093/mnras/sty2932
\bibitem[Fruchter et al.(2006)]{2006Natur.441..463F} Fruchter, A.~S., Levan, A.~J., Strolger, L., et al.\ 2006, \nat, 441, 463. doi:10.1038/nature04787
\bibitem[Gao \& Zhang(2015)]{2015ApJ...801..103G} Gao, H. \& Zhang, B.\ 2015, \apj, 801, 103. doi:10.1088/0004-637X/801/2/103
\bibitem[Gao et al.(2016)]{gao16} Gao, H., Zhang, B., L{\"u}, H.-J.\ 2016, \prd, 93, 044065 
\bibitem[Gehrels et al.(2005)]{2005Natur.437..851G} Gehrels, N., Sarazin, C.~L., O'Brien, P.~T., et al.\ 2005, \nat, 437, 851. doi:10.1038/nature04142
\bibitem[Giacomazzo et al.(2009)]{2009MNRAS.399L.164G} Giacomazzo, B., Rezzolla, L., \& Baiotti, L.\ 2009, \mnras, 399, L164. doi:10.1111/j.1745-3933.2009.00745.x
\bibitem[Gompertz et al.(2022)]{2022arXiv220505008G} Gompertz, B.~P., Ravasio, M.~E., Nicholl, M., et al.\ 2022, arXiv:2205.05008
\bibitem[Illarionov \& Sunyaev(1975)]{1975A&A....39..185I} Illarionov, A.~F. \& Sunyaev, R.~A.\ 1975, \aap, 39, 185
\bibitem[Kiuchi et al.(2014)]{2014PhRvD..90d1502K} Kiuchi, K., Kyutoku, K., Sekiguchi, Y., et al.\ 2014, \prd, 90, 041502. doi:10.1103/PhysRevD.90.041502
\bibitem[Kouveliotou et al.(1993)]{1993ApJ...413L.101K} Kouveliotou, C., Meegan, C.~A., Fishman, G.~J., et al.\ 1993, \apjl, 413, L101. doi:10.1086/186969
\bibitem[Lasky et al.(2014)]{lasky14} Lasky, P.~D., Haskell, B., Ravi, V., Howell, E.~J., \& Coward, D.~M.\ 2014, \prd, 89, 047302 
\bibitem[Lazarian et al.(2019)]{2019ApJ...882..184L} Lazarian, A., Zhang, B., \& Xu, S.\ 2019, \apj, 882, 184. doi:10.3847/1538-4357/ab2b38
\bibitem[Lee et al.(2000)]{2000PhR...325...83L} Lee, H.~K., Wijers, R.~A.~M.~J., \& Brown, G.~E.\ 2000, \physrep, 325, 83. doi:10.1016/S0370-1573(99)00084-8
\bibitem[{Lei et al.}(2005)]{Lei05} Lei, W. H., Wang, D. X., \& Ma, R. Y. 2005, ApJ, 619, 420
\bibitem[Lei et al.(2009)]{2009ApJ...700.1970L} Lei, W.~H., Wang, D.~X., Zhang, L., et al.\ 2009, \apj, 700, 1970. doi:10.1088/0004-637X/700/2/1970
\bibitem[Lei \& Zhang(2011)]{2011ApJ...740L..27L} Lei, W.-H. \& Zhang, B.\ 2011, \apjl, 740, L27. doi:10.1088/2041-8205/740/1/L27
\bibitem[Lei et al.(2013)]{2013ApJ...765..125L} Lei, W.-H., Zhang, B., \& Liang, E.-W.\ 2013, \apj, 765, 125. doi:10.1088/0004-637X/765/2/125
\bibitem[Lei et al.(2017)]{2017ApJ...849...47L} Lei, W.-H., Zhang, B., Wu, X.-F., et al.\ 2017, \apj, 849, 47. doi:10.3847/1538-4357/aa9074
\bibitem[Li(2000)]{2000PhRvD..61h4016L} Li, L.-X.\ 2000, \prd, 61, 084016. doi:10.1103/PhysRevD.61.084016
\bibitem[Liu et al.(2017)]{2017NewAR..79....1L} Liu, T., Gu, W.-M., \& Zhang, B.\ 2017, \nar, 79, 1. doi:10.1016/j.newar.2017.07.001
\bibitem[Liu et al.(2015)]{2015ApJS..218...12L} Liu, T., Hou, S.-J., Xue, L., et al.\ 2015, \apjs, 218, 12. doi:10.1088/0067-0049/218/1/12
\bibitem[Liu et al.(2012)]{2012ApJ...760...63L} Liu, T., Liang, E.-W., Gu, W.-M., et al.\ 2012, \apj, 760, 63. doi:10.1088/0004-637X/760/1/63
\bibitem[MacFadyen \& Woosley(1999)]{1999ApJ...524..262M} MacFadyen, A.~I. \& Woosley, S.~E.\ 1999, \apj, 524, 262. doi:10.1086/307790
\bibitem[Mangan et al.(2021)]{2021GCN.31210....1M} Mangan, J., Dunwoody, R., Meegan, C., et al.\ 2021, GRB Coordinates Network, Circular Service, No. 31210, 31210
\bibitem[Mei et al.(2022)]{2022arXiv220508566M} Mei, A., Banerjee, B., Oganesyan, G., et al.\ 2022, arXiv:2205.08566
\bibitem[{{Metzger} \& {Piro}(2014)}]{metzger14}
{Metzger}, B.~D., \& {Piro}, A.~L. 2014a, \mnras, 439, 3916
\bibitem[McKinney(2005)]{2005ApJ...630L...5M} McKinney, J.~C.\ 2005, \apjl, 630, L5. doi:10.1086/468184
\bibitem[Narayan et al.(1992)]{1992ApJ...395L..83N} Narayan, R., Paczynski, B., \& Piran, T.\ 1992, \apjl, 395, L83. doi:10.1086/186493
\bibitem[Narayan \& Yi(1994)]{1994ApJ...428L..13N} Narayan, R. \& Yi, I.\ 1994, \apjl, 428, L13. doi:10.1086/187381
\bibitem[Novikov \& Thorne(1973)]{1973blho.conf..343N} Novikov, I.~D. \& Thorne, K.~S.\ 1973, Black Holes (Les Astres Occlus), 343
\bibitem[Paczynski(1986)]{1986ApJ...308L..43P} Paczynski, B.\ 1986, \apjl, 308, L43. doi:10.1086/184740
\bibitem[Paczynski(1991)]{1991AcA....41..257P} Paczynski, B.\ 1991, \actaa, 41, 257
\bibitem[Paczy{\'n}ski(1998)]{1998ApJ...494L..45P} Paczy{\'n}ski, B.\ 1998, \apjl, 494, L45. doi:10.1086/311148
\bibitem[Proga \& Zhang(2006)]{2006MNRAS.370L..61P} Proga, D. \& Zhang, B.\ 2006, \mnras, 370, L61. doi:10.1111/j.1745-3933.2006.00189.x
\bibitem[Qin et al.(2013)]{2013ApJ...763...15Q} Qin, Y., Liang, E.-W., Liang, Y.-F., et al.\ 2013, \apj, 763, 15. doi:10.1088/0004-637X/763/1/15
\bibitem[Rastinejad et al.(2022)]{2022arXiv220410864R} Rastinejad, J.~C., Gompertz, B.~P., Levan, A.~J., et al.\ 2022, arXiv:2204.10864
\bibitem[{{Rezzolla} {et~al.}(2011){Rezzolla}, {Giacomazzo}, {Baiotti},
  {Granot}, {Kouveliotou}, \& {Aloy}}]{rezzolla11}
{Rezzolla}, L., {Giacomazzo}, B., {Baiotti}, L., {et~al.} 2011, \apjl, 732, L6
\bibitem[Riffert \& Herold(1995)]{1995ApJ...450..508R} Riffert, H. \& Herold, H.\ 1995, \apj, 450, 508. doi:10.1086/176161
\bibitem[Rosswog et al.(2014)]{rosswog14} Rosswog, S., Korobkin, O., Arcones, A., Thielemann, F.-K., \& Piran, T.\ 2014, \mnras, 439, 744 
\bibitem[Shakura \& Sunyaev(1973)]{1973A&A....24..337S} Shakura, N.~I. \& Sunyaev, R.~A.\ 1973, \aap, 24, 337
\bibitem[Shao \& Gao(2022)]{shao22} Shao, X. \& Gao, H.\ 2022, \apj, 927, 173. doi:10.3847/1538-4357/ac46a8
\bibitem[Song et al.(2018)]{2018MNRAS.477.2173S} Song, C.-Y., Liu, T., \& Li, A.\ 2018, \mnras, 477, 2173. doi:10.1093/mnras/sty783
\bibitem[Suvorov et al.(2022)]{2022arXiv220511112S} Suvorov, A.~G., Kuan, H.-J., \& Kokkotas, K.~D.\ 2022, arXiv:2205.11112
\bibitem[Uhm \& Zhang(2014)]{uhm14} Uhm, Z.~L. \& Zhang, B.\ 2014, Nature Physics, 10, 351. doi:10.1038/nphys2932
\bibitem[Wang et al.(2002)]{2002MNRAS.335..655W} Wang, D.~X., Xiao, K., \& Lei, W.~H.\ 2002, \mnras, 335, 655. doi:10.1046/j.1365-8711.2002.05652.x
\bibitem[Woosley(1993)]{1993ApJ...405..273W} Woosley, S.~E.\ 1993, \apj, 405, 273. doi:10.1086/172359
\bibitem[Woosley \& Bloom(2006)]{2006ARA&A..44..507W} Woosley, S.~E. \& Bloom, J.~S.\ 2006, \araa, 44, 507. doi:10.1146/annurev.astro.43.072103.150558
\bibitem[Xiao et al.(2022)]{2022arXiv220502186X} Xiao, S., Zhang, Y.-Q., Zhu, Z.-P., et al.\ 2022, arXiv:2205.02186
\bibitem[Yang et al.(2022)]{2022arXiv220412771Y} Yang, J., Zhang, B.-B., Ai, S., et al.\ 2022, arXiv:2204.12771
\bibitem[Yu et al.(2013)]{yu13} Yu, Y.-W., Zhang, B., \& Gao, H.\ 2013, \apjl, 776, L40 
\bibitem[Zhang \& Yan(2011)]{zhangyan11} Zhang, B. \& Yan, H.\ 2011, \apj, 726, 90. doi:10.1088/0004-637X/726/2/90
\bibitem[Zhang et al.(2018)]{2018MNRAS.475..266Z} Zhang, Q., Lei, W.~H., Zhang, B.~B., et al.\ 2018, \mnras, 475, 266. doi:10.1093/mnras/stx3229
\bibitem[Zhang et al.(2021)]{2021GCN.31236....1Z} Zhang, Y.~Q., Xiong, S.~L., Li, X.~B., et al.\ 2021, GRB Coordinates Network, Circular Service, No. 31236, 31236
\bibitem[Zhang et al.(2022)]{2022arXiv220509675Z} Zhang, H.-M., Huang, Y.-Y., Zheng, J.-H., et al.\ 2022, arXiv:2205.09675

\end{thebibliography}
\end{document}